\documentclass[superscriptaddress,twocolumn,pre]{revtex4}

\usepackage{amsmath}
\usepackage{graphicx,color}

\newcommand{\be}{\begin{equation}}
\newcommand{\ee}{\end{equation}}
\newcommand{\bea}{\begin{eqnarray}}
\newcommand{\eea}{\end{eqnarray}}

\begin{document}
\sloppy


\title{BEC dark matter, Zeldovich approximation and
   generalized Burgers  equation}

\author{Pierre-Henri Chavanis}
\affiliation{Laboratoire de Physique Th\'eorique (IRSAMC), CNRS and UPS, Universit\'e de Toulouse, F-31062 Toulouse, France}

\begin{abstract}
If the dark matter in the universe is a self-gravitating Bose-Einstein condensate (BEC) with quartic self-interaction described by the Gross-Pitaevskii-Poisson system, the adhesion model, the Burgers equation and the cosmological Kardar-Parisi-Zhang (KPZ) equation that have been introduced heuristically to solve the problems inherent to cold dark matter (CDM) models find a natural justification and an interesting generalization.
\end{abstract}

\maketitle


\section{Introduction}

The large-scale structure of the universe can be relatively well-explained by cold dark matter (CDM) models. In these models, the dark matter is represented by a collisionless gas without pressure (dust) interacting via Newtonian gravity only. It can be described by the Vlasov-Poisson system or, under additional approximations, by the pressureless Euler-Poisson system in an expanding background \cite{peeblesbook}. However, this model faces  several problems on the scale of galactic or sub-galactic structures. For example, CDM simulations lead to dark matter halos with density cusps \cite{nfw} while observations of rotation curves favor flat density profiles. In addition, it predicts an abundance of satellite galaxies around each galactic halo that is far beyond what we see around the Milky Way. At the cosmological level, nonlinear gravitational clustering is often studied by resorting to the  Zeldovich approximation \cite{zeldovich}. This approximation leads to the  inviscid Burgers equation which describes the free motion of fluid particles \cite{vergassola}. However, this equation develops shocks and formal singularities known as caustics (having the form of pancakes), so that it becomes invalid after particle-crossing. If we accept multi-stream solutions, the particles just cross each other after the caustic and pass through the pancakes instead of clustering into smaller objects like groups of galaxies. Therefore, a small-scale regularization must be introduced to overcome this problem and model the effects of ``punctuated'' gravitational attraction and pressure gradients that are not captured by the Zeldovich approximation.  Gurbatov {\it et al.} \cite{gurbatov} introduced the so-called ``adhesion model'' in which particles move according to the Zeldovich approximation until they fall into pancakes when their trajectories intersect, then ``stick'' to each other. This sticking can be modeled by introducing a viscosity in the Burgers equation in order to represent the effect of strong gravitational forces or pressure effects in the vicinity of a caustic.  Of course, the viscosity must be small in order to provide a smoothing effect at small-scales only (where particle-crossing occurs) but the limit $\nu\rightarrow 0$ is different from taking $\nu=0$. This model gives very good results in the nonlinear regime and can reproduce the skeleton of the ``cosmic web'' of large-scale structures (sheets, filaments, nodes) in $N$-body numerical simulations (see, e.g., Ref. \cite{ssmpm}).  A completely different  approach was developed  by Widrow \& Kaiser \cite{wk} who proposed to describe a classical collisionless self-gravitating gas by the  Schr\"odinger-Poisson system. In this approach, the constant $\hbar$ is not the Planck constant, but rather an adjustable parameter that controls the spatial resolution $\lambda_{deB}$ through a de Broglie relation $\lambda_{deB}=\hbar/mv$. It is argued that when $\hbar\rightarrow 0$, the Vlasov-Poisson system is recovered and that a finite value of $\hbar$ provides a small-scale regularization of the dynamics. In that case, the Schr\"odinger-Poisson system has nothing to do with quantum mechanics since it aims at describing the evolution of classical collisionless matter under the influence of gravity (in static or expanding universes). This model was further developed by Short \& Coles \cite{sc} who introduced a {\it free-particle approximation}. They showed that the dynamics of the particles before crossing is relatively close to the Zeldovich approximation but when crossing occurs the  quantum pressure  prevents the formation of singularities. Therefore, the quantum pressure  replaces the role of the viscosity in the adhesion model. Although these models have given interesting results and can be very useful in practice, one can argue, however, that their justification remains relatively {\it ad hoc}.

Recently, B\"ohmer \& Harko  \cite{bohmer} have proposed that dark matter could be a self-gravitating Bose-Einstein condensate (BEC) with quartic self-interaction described by the Gross-Pitaevskii-Poisson (GPP) system. This idea takes its origin in the concept of boson stars introduced by Kaup \cite{kaup} and Ruffini \& Bonazzola \cite{rb}. It is well-known that the Gross-Pitaevskii equation (or the nonlinear Schr\"odinger equation) can be reduced to hydrodynamical equations by means of the Madelung \cite{madelung} transformation. This yields the Euler-Poisson system with a ``classical'' isotropic pressure due to self-interaction and a quantum anisotropic pressure arising from the Heisenberg uncertainty principle. For a quartic self-interaction, the barotropic equation of state is that of a polytrope of index $n=1$. At large-scales, the pressure terms are negligible and one recovers the CDM model which has proven very successful. However, at small-scales, the classical and quantum  pressures can play a crucial role and regularize the dynamics by preventing the formation of singularities (caustics). BEC dark matter has very interesting properties: (i) the pressure can stabilize the system against gravitational collapse and lead to dark matter halos with flat density profiles. For very light bosons without self-interaction, the stabilization is due to the quantum pressure. For more massive bosons with self-interaction, the stabilization is due to repulsive scattering. Therefore, a BEC dark matter can solve the cusp problem and the missing satellite problem \cite{hu,bohmer,paper1}. (ii) BEC dark matter halos can reproduce the rotation curves of several low surface brightness galaxies \cite{arbey,bohmer}. (iii) At a cosmological level, a BEC dark matter can accelerate the formation of structures with respect to ordinary CDM models \cite{harko,cosmobec}. In this brief report, we make the additional remark that if dark matter is a self-gravitating BEC with quartic self-interaction, the adhesion approximation, the Burgers equation and the cosmological KPZ equation that have been introduced phenomenologically to solve the problems inherent to CDM models find a natural justification and an interesting generalization.

\section{Stochastic Gross-Pitaevskii-Poisson system}
\label{sec_gpp}

We assume that dark matter is a self-gravitating BEC with quartic self-interaction described by the  stochastic Gross-Pitaevskii-Poisson (GPP) system
\begin{equation}
\label{gpp1}
i\hbar \frac{\partial\psi}{\partial t}=-\frac{\hbar^2}{2m}\Delta\psi+m(\Phi+g Nm|\psi|^2+\eta)\psi,
\end{equation}
\begin{equation}
\label{gpp2}
\Delta\Phi=4\pi G Nm |\psi|^2,
\end{equation}
where $\psi({\bf r},t)$ is the wave function, $\rho({\bf r},t)=Nm|\psi|^2$ the density, $\Phi({\bf r},t)$ the gravitational
potential, $g={4\pi a_s\hbar^2}/{m^3}$ the pseudo-potential accounting for short-range interactions ($a_s$ is the s-scattering length) \cite{revuebec} and $\eta({\bf r},t)$ a stochastic potential (noise). We write the wave function in the form $\psi({\bf
r},t)=A({\bf r},t)e^{iS({\bf r},t)/\hbar}$ where $A$ and $S$ are real,
and make the Madelung \cite{madelung} transformation $\rho=Nm|\psi|^2=NmA^2$ and ${\bf u}=\nabla S/m$,
where $\rho({\bf r},t)$ is the density field and ${\bf u}({\bf r},t)$ the velocity field. We note that the flow is irrotational since $\nabla\times {\bf u}={\bf 0}$.  With this transformation, the stochastic GP equation (\ref{gpp1}) is {\it equivalent} to the stochastic barotropic Euler equations with an additional term $Q=-(\hbar^2/2m)\Delta\sqrt{\rho}/\sqrt{\rho}$ called the quantum potential (or quantum pressure). Indeed, one obtains the set of equations
\begin{equation}
\label{gpp4}
\frac{\partial\rho}{\partial t}+\nabla\cdot (\rho {\bf u})=0,\qquad \Delta\Phi=4\pi G \rho,
\end{equation}
\begin{equation}
\label{gpp5}
\frac{\partial {\bf u}}{\partial t}+({\bf u}\cdot \nabla){\bf u}=-g\nabla \rho-\nabla\Phi+\frac{\hbar^2}{2m^2}\nabla\left (\frac{\Delta \sqrt{\rho}}{\sqrt{\rho}}\right )-\nabla\eta.
\end{equation}
The first term on the r.h.s. of the Euler equation (\ref{gpp5}) can be interpreted as a classical isotropic pressure $-(1/\rho)\nabla p$ described by a barotropic equation of state $p=p(\rho)$. For a quartic self-interaction, it is given by
\begin{equation}
\label{gpp7}
p=\frac{1}{2}g\rho^2=\frac{2\pi a_s\hbar^2}{m^3}\rho^{2}.
\end{equation}
This is a polytropic equation of state of the form $p=K\rho^{\gamma}$
with polytropic index $n=1$ (i.e. $\gamma=1+1/n=2$) and polytropic constant $K=g/2={2\pi a_s\hbar^2}/{m^3}$. Other types of barotropic equations of state can be obtained depending on the form of the self-interaction. A detailed study of the time-independent solutions of the GPP system (or quantum barotropic Euler-Poisson system), connecting the non-interacting limit ($a_s\simeq 0$) \cite{rb} to the Thomas-Fermi (TF)  limit ($Q\simeq 0$) \cite{bohmer} has been given recently in Ref. \cite{paper1}.

\section{BEC equations in an expanding universe}
\label{sec_exp}

For the sake of simplicity, we consider an expanding Einstein-de Sitter (EdS) universe ($\Lambda=0$ and $\kappa=0$) \cite{peeblesbook} described by the equations
\begin{equation}
\label{exp1}
\rho_b a^3\sim 1,\qquad \ddot a=-\frac{4}{3}\pi G \rho_b a,\qquad {\dot a}^2=\frac{8}{3}\pi G\rho_b a^2,
\end{equation}
yielding $a\propto t^{2/3}$, $H=\dot a/{a}={2}/{3t}$ and $\rho_b={1}/{6\pi Gt^2}$, where  $\rho_b(t)$ is the background density,  $a(t)$ the scale factor and $H={\dot a}/a$ the Hubble constant. Our approach can be easily extended to more general models of  universe.

We shall first rewrite the hydrodynamic equations in the comoving frame. Making the change of variables ${\bf r}=a(t){\bf x}$, ${\bf u}=H{\bf r}+{\bf v}$ and $\phi=\Phi-\Phi_b$ where   ${\bf v}$ is the peculiar velocity and $\Phi_b=(2/3)\pi G\rho_b(t)r^2$ the background gravitational potential, and introducing the density contrast $\delta({\bf x},t)=\lbrack \rho({\bf x},t)-\rho_b(t)\rbrack/\rho_b(t)$, we obtain the system of equations \cite{cosmobec}:
\begin{equation}
\label{exp3}
\frac{\partial\delta}{\partial t}+\frac{1}{a}\nabla\cdot \lbrack (1+\delta) {\bf v}\rbrack=0,\qquad \Delta\phi=4\pi G \rho_b  a^2\delta,
\end{equation}
\begin{eqnarray}
\label{exp4}
\frac{\partial {\bf v}}{\partial t}+\frac{1}{a}({\bf v}\cdot \nabla){\bf v}+\frac{\dot a}{a}{\bf v}=-\frac{g \rho_b}{a}\nabla \delta\nonumber\\
-\frac{1}{a}\nabla\phi+\frac{\hbar^2}{2m^2a^3}\nabla \left (\frac{\Delta \sqrt{1+\delta}}{\sqrt{1+\delta}}\right )-\nabla\eta.
\end{eqnarray}
For simplicity, we shall always denote the noise by $\eta$ although a new notation should be introduced after each transformation. At large scales, pressure  and noise effects are negligible and the CDM model ($\hbar=p=\eta=0$) is recovered. However, pressure effects become important when nonlinear structures form, and the BEC dark matter model (\ref{exp3})-(\ref{exp4}) should be used instead.

Measuring the evolution in terms of $a$ rather than in terms of $t$ and introducing the new velocity ${\bf w}={\bf v}/a\dot a$ and the new gravitational potential $\psi=\phi/4\pi G\rho_b a^3$, we can recast the foregoing equations in the form
\begin{equation}
\label{exp6}
\frac{\partial\delta}{\partial a}+\nabla\cdot \lbrack (1+\delta) {\bf w}\rbrack=0, \qquad \Delta\psi=\frac{\delta}{a},
\end{equation}
\begin{eqnarray}
\label{exp7}
\frac{\partial {\bf w}}{\partial a}+({\bf w}\cdot \nabla){\bf w}+\frac{3}{2a}({\bf w}+\nabla\psi)=-\frac{3 a_s\hbar^2}{2Gm^3a^4}\nabla \delta\nonumber\\
+\frac{3\hbar^2}{16\pi G m^2\rho_b a^6}\nabla \left (\frac{\Delta \sqrt{1+\delta}}{\sqrt{1+\delta}}\right )-\nabla\eta.
\end{eqnarray}

For short times, the perturbations are small $\delta\ll 1$, $\phi\ll 1$, $|{\bf w}|\ll 1$, and the hydrodynamic equations can be linearized.  We obtain
\begin{equation}
\label{lin1}
\frac{\partial\delta}{\partial a}+\nabla\cdot  {\bf w}=0,\qquad \Delta\psi=\frac{\delta}{a},
\end{equation}
\begin{eqnarray}
\label{lin2}
\frac{\partial {\bf w}}{\partial a}+\frac{3}{2a}({\bf w}+\nabla\psi)=-\frac{3 a_s\hbar^2}{2Gm^3a^4}\nabla \delta\nonumber\\
+\frac{3\hbar^2}{32\pi G m^2\rho_b a^6}\nabla(\Delta \delta)-\nabla\eta.
\end{eqnarray}
Taking the ``time'' derivative of Eq. (\ref{lin1}-a) and the divergence of Eq. (\ref{lin2}), these equations  can be combined into a single equation for the density contrast
\begin{eqnarray}
\label{lin3}
\frac{\partial^2\delta}{\partial a^2}+\frac{3}{2a}\frac{\partial\delta}{\partial a}=\frac{3\delta}{2a^2}\nonumber\\
+\frac{3a_s\hbar^2}{2Gm^3a^4}\Delta\delta-\frac{3\hbar^2}{32\pi G m^2\rho_b a^6}\Delta^2\delta+\Delta\eta.
\end{eqnarray}
This equation has been studied in Ref. \cite{cosmobec} in the no-noise limit.
At large scales, we can ignore pressure and noise effects, and we recover the equation for the density contrast of a cold gas
\begin{eqnarray}
\label{lin4}
\frac{\partial^2\delta}{\partial a^2}+\frac{3}{2a}\frac{\partial\delta}{\partial a}=\frac{3\delta}{2a^2}.
\end{eqnarray}
The growing solution to this equation is $\delta_+({\bf x},a)=aD({\bf x})$ \cite{peeblesbook}. Then, Eq. (\ref{lin1}-b) implies that $\psi({\bf x},a)=\psi({\bf x})$ is constant. Therefore, $\delta_+({\bf x},a)=a\Delta\psi({\bf x})$. On the other hand, in the cold gas approximation, Eq. (\ref{lin2}) reduces to
\begin{eqnarray}
\label{lin5}
\frac{\partial {\bf w}}{\partial a}+\frac{3}{2a}({\bf w}+\nabla\psi)={\bf 0}.
\end{eqnarray}
After a transient regime, the velocity field tends toward  the solution ${\bf w}({\bf x},a)=-\nabla\psi({\bf x})$.

\section{Zeldovich approximation and a generalized equation of structure formation}
\label{sec_z}

The Zeldovich approximation \cite{zeldovich} amounts to extending this relation to the (weakly) nonlinear regime, i.e.  ${\bf w}({\bf x},a)\simeq -\nabla\psi({\bf x},a)$. The fact that ${\bf w}$ is a potential flow in the BEC model gives further support to this approximation. In that case, Eq. (\ref{exp7}) reduces to the form
\begin{eqnarray}
\label{z1}
\frac{\partial {\bf w}}{\partial a}+({\bf w}\cdot \nabla){\bf w}=-\frac{3 a_s\hbar^2}{2Gm^3a^4}\nabla \delta\nonumber\\
+\frac{3\hbar^2}{16\pi Gm^2\rho_b a^6}\nabla \left (\frac{\Delta \sqrt{1+\delta}}{\sqrt{1+\delta}}\right )-\nabla\eta,
\end{eqnarray}
where the explicit dependence on the gravitational potential has disappeared. Since $\delta=a\Delta\psi=-a\nabla\cdot {\bf w}$ and $\nabla\delta=-a\Delta {\bf w}$, we can rewrite the foregoing equation as
\begin{eqnarray}
\label{z2}
\frac{\partial {\bf w}}{\partial a}+({\bf w}\cdot \nabla){\bf w}=\frac{3 a_s\hbar^2}{2Gm^3a^3}\Delta{\bf w}
\nonumber\\
+\frac{3\hbar^2}{16\pi Gm^2\rho_b a^6}\nabla \left (\frac{\Delta \sqrt{1-a\nabla\cdot {\bf w}}}{\sqrt{1-a\nabla\cdot {\bf w}}}\right )-\nabla\eta.
\end{eqnarray}
This can be viewed as a generalized noisy Burgers equation with an additional quantum pressure term.  Since ${\bf w}$ is a potential flow, we obtain
\begin{eqnarray}
\label{z3}
\frac{\partial \psi}{\partial a}=\frac{(\nabla\psi)^2}{2}+\frac{3 a_s\hbar^2}{2Gm^3a^3}\Delta\psi\nonumber\\
-\frac{3\hbar^2}{16\pi Gm^2\rho_b a^6} \frac{\Delta \sqrt{1+a\Delta\psi}}{\sqrt{1+a\Delta\psi}}+\eta.
\end{eqnarray}
This can be viewed as a cosmological KPZ equation \cite{kpz}. Without the forcing term, this is just the Bernouilli equation, or the Hamilton-Jacobi equation, with an additional quantum potential. We note that the potential $\psi$ is related to the phase $S$ of the wave function of the BEC by
$\psi=-(1/a)(S/ma^2H-x^2/2)$. A noisy Burgers equation and a KPZ equation have been introduced and studied previously in cosmology by Buchert {\it et al.} \cite{buchert}, Jones \cite{jones}, Coles \cite{coles} and Matarrese \& Mohayaee \cite{mm}.

Let us consider particular cases of Eq. (\ref{z2}).

(i) For $a_s=\hbar=\eta=0$, we get the inviscid Burgers equation which is equivalent  to the Zeldovich approximation.

(ii) If we neglect the quantum potential, we get the noisy Burgers equation
\begin{eqnarray}
\label{z5}
\frac{\partial {\bf w}}{\partial a}+({\bf w}\cdot \nabla){\bf w}=\nu(a)\Delta{\bf w}-\nabla\eta,
\end{eqnarray}
with a time-dependent viscosity given by $\nu(a)={3 a_s\hbar^2}/{2Gm^3a^3}$ (this equation is time-reversible via $a\rightarrow -a$ and ${\bf w}\rightarrow -{\bf w}$). 
Therefore, the BEC dark matter hypothesis leads to a natural justification of the (stochastic) adhesion model. Another justification was previously given by Buchert {\it et al.} \cite{buchert} in terms of a coarse-graining process inherent to a hydrodynamic description. However, to arrive at the Burgers equation, several simplifying hypothesis had to be introduced: (i) dark matter is described by hydrodynamic equations with an {\it isotropic} pressure; (ii) the equation of state is a polytrope $p=K\rho^{\gamma}$ with $\gamma=2$; (iii) the velocity field and the stochastic force derive from a potential. Interestingly, these properties directly result from the BEC dark matter hypothesis (i.e. from the GP equation) without further assumption. Note also that the change of variable $\psi({\bf x},a)=2\nu(a)\ln W({\bf x},a)$ transforms Eq. (\ref{z3}) into
\begin{eqnarray}
\label{z3b}
\frac{\partial W}{\partial a}=\nu(a)\Delta W+\left\lbrack \frac{1}{2\nu}\eta({\bf x},a)+\frac{3}{a}\ln W\right \rbrack W. 
\end{eqnarray}

(iii) In the absence of self-interaction and noise, we obtain an equation of the form
\begin{eqnarray}
\label{z7}
\frac{\partial {\bf w}}{\partial a}+({\bf w}\cdot \nabla){\bf w}=\frac{3\hbar^2}{16\pi Gm^2\rho_b a^6}\nabla \left (\frac{\Delta \sqrt{1-a\nabla\cdot {\bf w}}}{\sqrt{1-a\nabla\cdot {\bf w}}}\right ).\nonumber\\
\end{eqnarray}
This can be viewed as a Burgers equation where the small-scale regularization is provided by the quantum pressure. Our approach has some  similarities with the effective wave mechanics approach of Short \& Cole \cite{sc}. However, in our approach, the coefficient in front of the quantum potential in Eq. (\ref{z7}) depends on ``time'' $a$, while in the approach of Short \& Cole \cite{sc}, this coefficient is constant (see their Eq. (24)). This is because they introduce an effective Schr\"odinger equation [their Eq. (23)] directly in the comoving frame while we start from the (nonlinear) Schr\"odinger equation (\ref{gpp1}) in the inertial frame. When we write the (nonlinear) Schr\"odinger equation (\ref{gpp1}) in the comoving frame, we obtain an equation different from their Eq. (23). Ribeiro \& Peixoto de Faria \cite{rp} have  also developed an effective wave mechanics approach in which they relate the gradient of the quantum potential to the Laplacian of a kinematical velocity plus a noise term. Their approach, which is essentially phenomenological, leads to a result different from Eq. (\ref{z7}).

\section{Conclusion}
\label{sec_conclusion}

In this brief report, we have shown that the assumption that dark matter is a self-gravitating Bose-Einstein condensate with quartic self-interaction described by the (stochastic) GPP system leads to a natural justification of the adhesion model, the Burgers equation and the cosmological KPZ equation without more hypothesis than the Zeldovich approximation that is common to most works on the subject. Therefore, not only the BEC model is consistent with previous works, but it generalizes them and extends their scope. In addition, it gives a new justification of these phenomenological models. This result adds to the other nice properties of BEC dark matter (flat density profiles, flat rotation curves, acceleration of the growth of perturbations...) recalled in the Introduction.

The BEC model not only follows the general evolution of inhomogeneities but it also describes the internal structure of density enhancements. In this sense, it improves upon the standard adhesion model.  Indeed, at the level of dark matter halos, the equations reduce to the condition of hydrostatic equilibrium and lead to virialized structures similar to $n=1$ polytropes (other configurations could be obtained depending on the form of self-interaction and on the equation of state). Therefore, BEC dark matter provides a model that describes both the large-scale structures of the universe (through the generalized Burgers equation (\ref{z2})) and the structure of dark matter halos (through the explicit expression of the pressure (\ref{gpp7}) arising from short-range interactions).

If we justify the GPP system in terms of quantum mechanics, we must assume that the mass $m$ of the bosons is extraordinarily small ($m< 10^{-24}\, {\rm eV}/c^2$!) for quantum mechanics to be relevant on galactic scales. This has been called ``fuzzy dark matter'' in Ref. \cite{hu}. Alternatively,  one can produce similar quantum effects with larger boson masses if the particles have a self-interaction \cite{colpi}. Since the nature of dark matter is not known, we cannot reject {\it a priori} that quantum mechanics plays some role at galactic or cosmological scales. In any case, the GPP system (\ref{gpp1})-(\ref{gpp2}) can always be viewed as an effective wave mechanics approach with an adjustable Planck constant which generalizes the initial model of  Widrow \& Kaiser \cite{wk} based on the Schr\"odinger-Poisson (SP) system. Finally, other approaches to the problem, based on the Vlasov-Poisson (VP) system, could be contemplated, see e.g. Ref. \cite{csr}.

\end{document}